\newcommand\percc{\ifmmode{\rm cm^{-3}}\else{cm$^{-3}$}\fi}
\newcommand\kms{\km\,\pers}
\newcommand\cmsq{\ifmmode{\rm cm^2}\else{cm$^2$}\fi}
\newcommand\coldv{\ifmmode{\cmtwo\,\km^{-1}\,{\rm s}} 
	\else {\cmtwo\,\km$^{-1}\,$s}\fi}
\newcommand\cmtwo{\ifmmode{\rm cm^{-2}}\else{cm$^{-2}$}\fi}
\newcommand\cm{\ifmmode{\rm cm}\else{cm}\fi}
\newcommand\km{\ifmmode{\rm km}\else{km}\fi}
\newcommand\hz{\ifmmode{\rm Hz}\else{Hz}\fi}
\newcommand\mhz{\ifmmode{\rm MHz}\else{MHz}\fi}
\newcommand\ghz{\ifmmode{\rm GHz}\else{GHz}\fi}
\newcommand\pers{\ifmmode{\rm s^{-1}}\else{s$^{-1}$}\fi}
\newcommand\kkms{\ifmmode{\rm K \kms}\else{K \kms}\fi}
\newcommand\pdeg{\fdg}
\newcommand\pmin{\farcm}
\newcommand\psec{\farcs}

\newcommand\gsim{\ga}
\newcommand\lsim{\la}
\def\ee#1{\ifmmode{\times 10^{#1}}\else{$\times 10^{#1}$}\fi}
\newcommand\jon{\ifmmode{J\!=\!1\!\rightarrow\!0}\else{$J\!=\!1\!\rightarrow\!0$}\fi}
\newcommand\jtw{\ifmmode{J\!=\!2\!\rightarrow\!1}\else{$J\!=\!2\!\rightarrow\!1$}\fi}
\newcommand\jth{\ifmmode{J\!=\!3\!\rightarrow\!2}\else{$J\!=\!3\!\rightarrow\!2$}\fi}
\newcommand\jfo{\ifmmode{J\!=\!4\!\rightarrow\!3}\else{$J\!=\!4\!\rightarrow\!3$}\fi}
\newcommand\jfi{\ifmmode{J\!=\!5\!\rightarrow\!4}\else{$J\!=\!5\!\rightarrow\!4$}\fi}
\newcommand\jsi{\ifmmode{J\!=\!6\!\rightarrow\!5}\else{$J\!=\!6\!\rightarrow\!5$}\fi}
\newcommand\jse{\ifmmode{J\!=\!7\!\rightarrow\!6}\else{$J\!=\!7\!\rightarrow\!6$}\fi}
\newcommand\jei{\ifmmode{J\!=\!8\!\rightarrow\!7}\else{$J\!=\!8\!\rightarrow\!7$}\fi}
\newcommand\maff{\ifmmode{\rm Maffei~2}\else{Maffei~2}\fi}
\newcommand\msun{\ifmmode{M_\odot}\else{$M_\odot$}\fi}
\newcommand\rsun{\ifmmode{R_\odot}\else{$R_\odot$}\fi}
\newcommand\lsun{\ifmmode{L_\odot}\else{$L_\odot$}\fi}
\newcommand\lfir{\ifmmode{L_{FIR}}\else{$L_{FIR}$}\fi}
\newcommand\vlsr{\ifmmode{V_{LSR}}\else{$V_{LSR}$}\fi}
\newcommand\tmb{\ifmmode{T_{mb}}\else{$T_{mb}$}\fi}
\newcommand\tx{\ifmmode{T_{ex}}\else{$T_{ex}$}\fi}
\newcommand\tdv{\ifmmode{\int\nolimits\tmb dv}\else{$\int\nolimits\tmb dv$}\fi}
\newcommand\tas{\ifmmode{T_A^*}\else{$T_A^*$}\fi}
\newcommand\tsys{\ifmmode{T_{sys}}\else{$T_{sys}$}\fi}
\newcommand\nhh{\ifmmode{n_\htwo}\else{$n_\htwo$}\fi}
\newcommand\ncrit{\ifmmode{n_{crit}}\else{$n_{crit}$}\fi}
\newcommand\rat{\ifmmode{\cal R}\else{${\cal R}$}\fi}

\newcommand\gray{\ifmmode{\gamma{\rm -ray}}\else{$\gamma$-ray}\fi}
\newcommand\grays{\ifmmode{\gamma{\rm -rays}}\else{$\gamma$-rays}\fi}

\newcommand\hi{H~{\small I}}

\newcommand\ha{\ifmmode{{\rm H}\alpha}\else{H$\alpha$}\fi}
\newcommand\htwo{\ifmmode{{\rm H}_2}\else{H$_2$}\fi}
\newcommand\hcop{\ifmmode{\rm HCO}^+\else{HCO$^+$}\fi}
\newcommand\ihcop{\ifmmode{\rm H^{13}CO^+}\else{H$^{13}$CO$^+$}\fi}
\newcommand\ihcn{\ifmmode{\rm H^{13}CN}\else{H$^{13}$CN}\fi}
\newcommand\ico{\ifmmode{\rm ^{13}CO}\else{$^{13}$CO}\fi}
\newcommand\co{\ifmmode{\rm ^{12}CO}\else{$^{12}$CO}\fi}
\newcommand\ics{\ifmmode{\rm C^{34}S}\else{C$^{34}$S}\fi}

\documentclass[12pt,preprint]{aastex}

\begin{document}

\title{The Structure, Kinematics and Physical Properties of the
Molecular Gas in the Starburst Nucleus of NGC 253}

\author{Timothy A. D. Paglione$^1$, Omar Yam$^2$, Tomoka Tosaki$^3$,
James M. Jackson$^4$}
\email{paglione@york.cuny.edu}
\altaffiltext{1}{York College, City University of New York, 94-20 Guy
R. Brewer Blvd., Jamaica, NY 11451}
\altaffiltext{2}{Instituto Nacional de Astrof\'{\i}sica, Optica y
Electr\'onica, Apartado Postal 216 y 51, 72000 Puebla, Pue., M\'exico}
\altaffiltext{3}{Gunma Astronomical Observatory, 6860-86 Nakayama,
Takayama, Agatsuma, Gunma 377-0702, Japan}
\altaffiltext{4}{Department of Astronomy, Boston University, 725
Commonwealth Ave., Boston, Massachusetts 02215}

\begin{abstract}

We present $5\psec2\times 2\psec6$ resolution interferometry of CO
\jon\ emission from the starburst galaxy NGC 253.  The high spatial
resolution of these new data, in combination with recent high
resolution maps of \ico, HCN and near-infrared emission, allow us for
the first time to link unambiguously the gas properties in the central
starburst of NGC 253 with its bar dynamics.  We confirm that the star
formation results from bar-driven gas flows as seen in ``twin peaks''
galaxies.  Two distinct kinematic features are evident from the CO map
and position-velocity diagram: a group of clouds rotating as a solid
body about the kinematic center of the galaxy, and a more extended gas
component associated with the near-infrared bar.  We model the line
intensities of CO, HCN and \ico\ to infer the physical conditions of
the gas in the nucleus of NGC 253.  The results indicate increased
volume densities around the radio nucleus in a twin-peaks morphology.
Compared with the CO kinematics, the gas densities appear highest near
the radius of a likely inner Linblad resonance, and slightly lead the
bar minor axis.  This result is similar to observations of the
face-on, twin-peaks galaxy NGC 6951, and is consistent with models of
starburst generation due to gas inflow along a bar.

\end{abstract}


\section{Introduction}

Bar-driven gas flows have been proposed to explain the high star
formation rates in the nuclei of some starburst galaxies
\citep*{twinpeaks, knapen, kohno, sheth}.  Star formation is predicted 
to be enhanced in regions with little shear and/or weak or no shocks,
i.e., where the velocity gradients are small.  These regions are
predicted to be found, and in some cases are observed, at the bar
ends, along the leading edges of the bar, and at locations within a
nuclear ring or spiral.  Molecular gas concentrations often appear in
these areas, in particular at so-called ``twin peaks'' near the
contact points between the dust lanes of the bar and the nuclear ring
\citep{twinpeaks}.

High resolution images of the molecular gas in the archetypal
starburst galaxy NGC 253 suggest that a similar ring of clouds may
exist around its nucleus.  Specifically, the position-velocity diagram
(PVD) shows a hole near the center \citep{n253.hcop, n253.cs,
n253.ihcop}, though this feature does not unambiguously indicate the
presence of a ring \citep{sakamoto}.  From CS observations,
\citet{n253.cs} suggest that much of the dense, star-forming gas in
the nucleus of NGC 253 lies in a ring associated with so-called $x_2$
orbits, which are small, elongated orbits perpendicular to the bar
\citep{bt}.  Similarly, \citet{n253.ihcop} conclude from the PVDs of
\ihcop\ and SiO emission in NGC 253 that the dense gas also follows
these orbits.  The ratio of HCN and CO intensities, an indicator of
density, is also elevated 5--$10''$ on either side of the nucleus
along the major axis \citep*{n253.hcn}.  Indeed, dense gas in the
face-on galaxy NGC 6951 appears to follow the $x_2$ orbits closely
\citep{kohno}, but the low declination and high inclination of NGC 253
\citep[$87\pdeg5$,][]{pence} have hindered similar analysis.

Previous high resolution maps of gas emission from NGC 253 either did
not trace the bulk of the molecular gas, i. e., they were most
sensitive to dense, shocked or ionized gas \citep{n253.hcn, n253.h92a,
n253.cs, n253.ihcop}, or they lacked the spatial resolution to isolate
individual cloud complexes and their kinematics \citep*{canzian,
n253.co21}.  Further, since the first CO interferometer map of NGC 253
by \citet{canzian}, new images have been made of the stellar and dust
distributions \citep{2mass} and optically thin gas traced by \ico\
\citep{n253.ico}.  \citet{das} have also analyzed the CO velocity
field at high resolution, and with the Submillimeter Array now 
operational, interferometric maps of dense and warm gas traced by CO
\jth\ and other lines are certainly imminent.  Given these new and
anticipated data, it is appropriate to reexamine the properties of the
molecular gas in NGC 253 but on the spatial scales of individual giant
molecular clouds ($\lsim 40$ pc).  To this end, we imaged the CO \jon\
emission from NGC 253 with the Nobeyama Millemeter Array.  To place
any new work in the proper context will require this relied-upon
tracer of the bulk of the star-forming gas in galaxies.  We compare
these data to high resolution HCN and \ico\ maps \citep{n253.hcn,
n253.ico} to infer the properties of the molecular clouds and compare
them to the kinematics and dust distribution to test how the bar in
NGC 253 may drive the central starburst.

\section{Observations}

The CO \jon\ data were obtained 1996 January and February from the
6-element Nobeyama Millimeter Array at the Nobeyama Radio
Observatory (NRO)\footnote{The NRO is a branch of the National
Astronomical Observatory, operated by the Ministry of Education,
Culture, Sports, Science and Technology, Japan.} using the AB and C
antenna configurations.  The channels were averaged to achieve a
velocity resolution of roughly 10 \kms.  The data were gridded with
natural weighting, mapped, and CLEANed with the NRAO AIPS package.
The FWHM of the synthesized beam is $5\psec2\times 2\psec6$, with a
position angle (PA) of $1\pdeg5$.  The total integrated CO \jon\ flux
smoothed to 43$''$ is 3670 Jy beam$^{-1}$ \kms, roughly 50\%\ of the
single-dish integrated intensity \citep{n253.co32, fcrao.cosurvey,
sorai, umigs}.

We recovered a 2.6 mm continuum map by averaging the line-free
channels in the data cube.  We also created a continuum image from a
linear baseline subtraction within the datacube.  Both methods
produced nearly identical continuum maps.  The emission appears
unresolved, but may be marginally resolved only along the major axis
of the galaxy.  The peak flux is $120\pm 30$ mJy, which is consistent
with similar measurements \citep{n253.cs, n253.ihcop}.

The CO map originally appeared to have a registration offset from
previous work.  The large angular distance of $12\degr$ (mostly north)
to the phase calibrator 0050-094 may be the cause, though no strong
phase variations were found during the observations.  To properly
compare this map to previous work, we aligned the 2.6 mm continuum
source with the nuclear radio source of \citet{ua}.  A total shift
(mostly north) of roughly $9''$  was required.  A subsequent minor
adjustment in declination was made to align the major axis and
velocity centroid ($\vlsr = 235 \kms$) of the CO emission with those
of the high resolution HCN map.  This adjustment was a small fraction
of the beam width in declination.  In summary, the central velocity
field, which is roughly perpendicular to the major axis, constrains
the R.A. offset, and the declination offset is mostly constrained by
the cloud distribution along the major axis.  Judging from the
alignment of the continuum sources, major axes and velocity fields, we
estimate that the CO map is aligned to within half a beam width in
R.A. and Decl. ($1\psec3$ and $2\psec6$, respectively; the beam is
oriented nearly north-south) with the radio continuum, HCN and \ico\
maps.

\section{Results}

\begin{figure}
\includegraphics[angle=90,scale=.75]{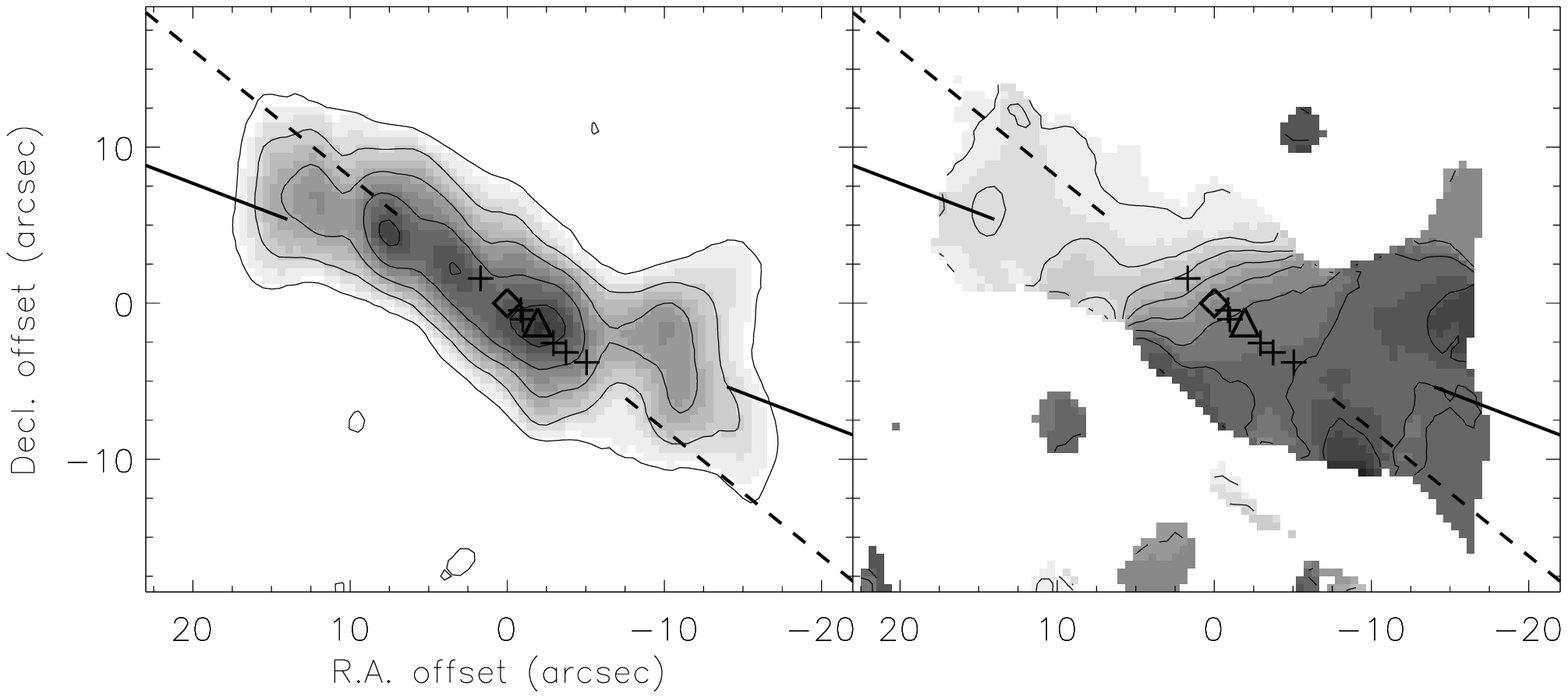}
\caption{(Left) Integrated intensity map of CO \jon\ emission from NGC
253.  The contours are 10, 30, 50, 70, and 90\%\ of the peak, 2118
\kkms\ (315 Jy beam$^{-1}$ \kms).  (Right) Velocity field of the CO
emission.  The contours are, from white to black, 125 to 300 \kms\ by
25 \kms.  The galaxy's major axis (dashed line) and bar axis (solid
line) are shown.  The locations of the thermal radio sources
(crosses), 1.3 cm peak \citep[diamond,][]{ua}, and mid-infrared peak
\citep[triangle,][]{keto} are indicated.
\label{fig.moms}}
\end{figure}

\begin{figure}
\includegraphics[angle=-90,scale=.65]{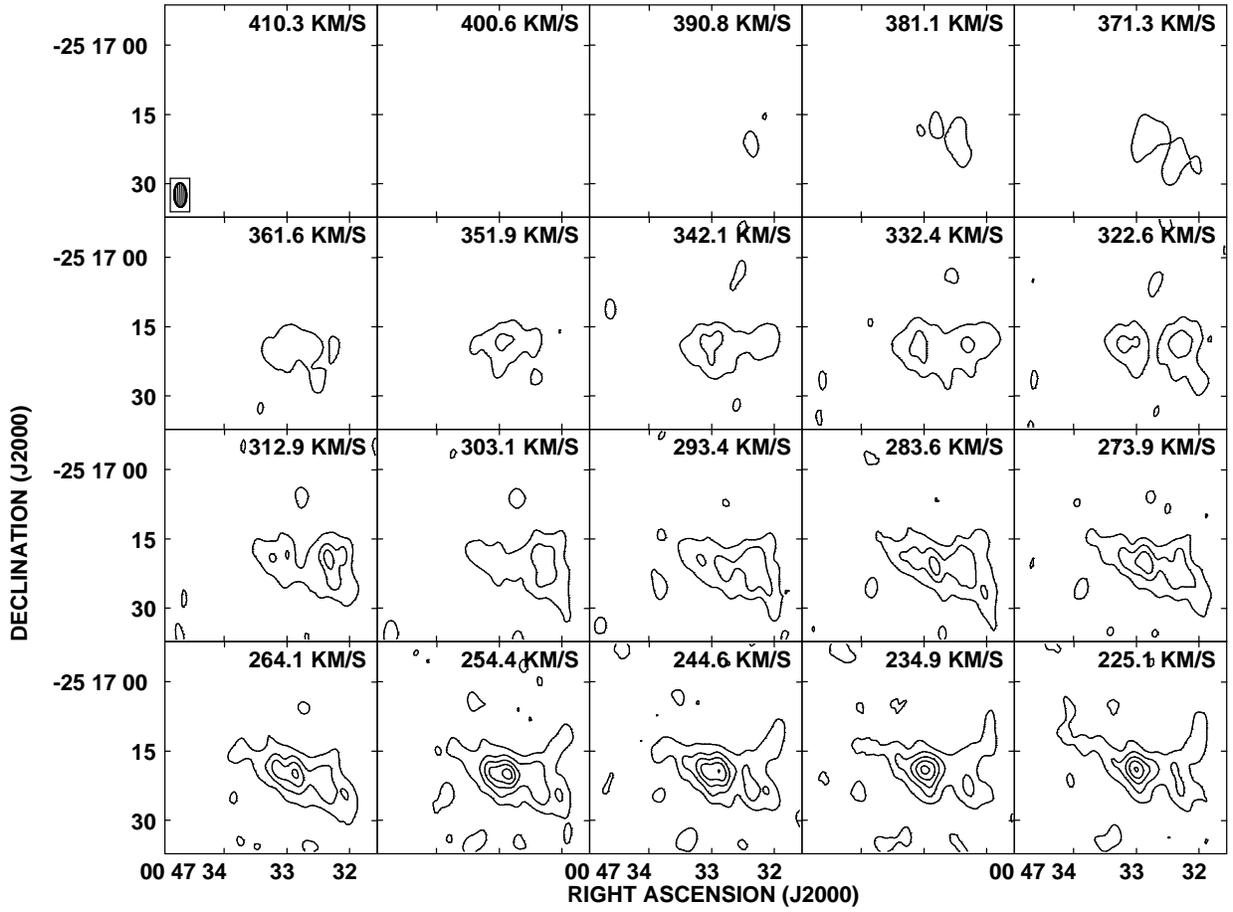}
\caption{Channel maps of CO emission in NGC 253.  The contours are 10,
30, 50, 70, and 90\%\ of the peak, 22.2 K (3.3 Jy
beam$^{-1}$).
\label{fig.chmaps}}
\end{figure}

Maps of the integrated CO intensity and velocity field are shown in
Figure~\ref{fig.moms}.  Figure~\ref{fig.chmaps} shows the CO channel
maps.  The general morphology and velocity field are consistent with
other high resolution maps \citep{canzian, n253.hcn, n253.co21,
n253.cs, n253.ihcop, das}.  There is an extended ridge of emission
aligned with the infrared bar at a PA of $69\degr$, and a central
group of bright clouds aligned with the major axis of the galaxy
(PA = $51\degr$, \citet{pence}) orbiting the radio nucleus in a
possible ring.  The kinematics of these features are discussed below.

The CO emission is somewhat more extended than the maps of dense gas
tracers such as HCN and \ihcop, with intense emission at the radio
nucleus and the mid-infrared peak \citep{keto}.  The CO peak appears
to lie between the thermal radio sources, unlike the HCN emission
which at the same resolution is strongly coincident with the thermal
sources.  The CO spectra agree well with the $16''$ resolution data of
\citet{sorai} and the $12''$ resolution CO \jtw\ data
\citep{n253.co21}.  We even recover many double-peaked spectra as seen
at $16''$ resolution.  The line widths of the new data are slightly
narrower than, and the peak line temperatures higher than previous
work, indicating that we are better resolving the star-forming clouds.
The overall CO \jon\ and \jtw\ distributions and PVDs at similar
resolution are comparable, though the new map resolves the emission
for the first time into small-scale complexes.  Also for the first
time at these spatial scales, the ratio of the CO \jtw\ and \jon\
integrated intensities is found to be $\gsim 1$ over much of the map,
implying optically thick gas.  The gas properties are discussed
below.

The CO emission is compared to the 2MASS $K_s$ and $H-K_s$ maps
\citep{2mass} in Figure~\ref{fig.2mass}.  The $K_s$ map clearly shows
the bar and some of the ring structure around the nucleus.  The
wide-field, $16''$ resolution map of \citet{sorai} indicates that the
CO closely traces the infrared bar distribution.  The new CO map shows
that the small-scale CO distribution in the central arcminute matches
up with the dust in the bar.  The bar appears to have leading dust
ridges that meet up with the nuclear CO emission.  This morphology is
typical for twin-peaks galaxies.

The extended ridge of emission and central cloud distribution have
distinct velocity structures according to the CO velocity field.  The
central clouds move around the radio nucleus as a solid body with a
PA of $\sim 55\degr$ (Figure~\ref{fig.moms}), similar to the
distribution of the thermal radio sources, central HCN clouds and
central near-infrared emission \citep{n253.hcn, ua, 2mass}.  However,
unlike what is expected for solid-body rotation, the velocity field of
these clouds is not exactly perpendicular to the cloud distribution.
This result was seen by \citet{das} as well, and may result from
unresolved, complex gas motions at the nucleus.  The gas kinematics
are discussed more extensively in Section~\ref{kin}.

The peak CO flux is quite high, $3.3 \pm 0.1$ mJy beam$^{-1}$,
corresponding to a line brightness temperature of $22.2 \pm 0.8$ K.
Given a likely beam filling factor below unity implies hot clouds ($>
25$ K) in the nucleus of NGC 253, consistent with single-dish studies
of higher frequency CO transitions \citep*{harrison, n253.co76}.

\section{Gas Properties}

To estimate the physical properties of the clouds in NGC 253, we
compare the CO data with maps of HCN and \ico\ at similar resolution
\citep{n253.hcn, n253.ico}, and model the emission.  The model assumes
the emission originates in unresolved, homogeneous, spherical clouds.
A photon escape probability function is included to account for the
radiative excitation of optically thick lines \citep{rt}.  We model
the emission from the first 11 levels of CO and first 8 levels of HCN.
The collision rate coefficients are from the literature (CO: Flower \&
Launay 1985; HCN: Green \& Thaddeus 1974).  We vary the [CO]/[\ico]
abundance ratio from 30, appropriate for Galactic center clouds
\citep{lp}, to 50, which was derived from a lower limit to the
CN/$^{13}$CN intensity ratio in NGC 253 \citep{n253.icn}.

To take advantage of the highest resolution data, we begin by
modeling only the ratio of integrated intensities of HCN and CO,
which we refer to as simply HCN/CO.  To this analysis we then include
the somewhat lower resolution \ico\ data to further constrain the
results (Figure~\ref{fig.rats}).  With each new modeled species, we
assume relative abundances and degrade the resolution on the map, but
the relative cloud properties, i.e., gradients over 40--100 pc scales,
are well constrained in the end.  All maps were aligned by the
emission distributions on the major axis, available continuum
positions, and velocity fields as described above.

The HCN/CO ratio is sensitive to the volume density, but the kinetic
temperature, $T_k$ and CO column density are not well constrained.  We
estimate $T_k$ iteratively as in \citet{hcnco} using the CO line
brightness temperature, such that $T_k = T_{CO}/f_1 + 2.725$ K.  The
factor $f_1 \le \phi (1-e^{-\tau})$ accounts for non-thermal
excitation, beam dilution and opacity.  The emission is initially
assumed to be thermalized and optically thick ($f_1 = \phi$, the area
beam filling factor).  The column density is initially estimated by
assuming a constant conversion factor between the CO integrated
intensity and the molecular gas column density, $X = 0.5\times10^{20}$
\cmtwo (\kkms)$^{-1}$ \citep{sorai, umigs}, and a CO abundance.  This
beam-averaged column density is converted to a cloud column density
with the beam filling factor.  The final beam filling factor is found
by comparing the observed CO temperature with the value predicted by
the model.  Obviously the choice of $f_1$ affects the resultant
columns and temperatures.  The relative HCN and CO abundances are
varied for different model runs, but remain within the range of
predicted and observed values for dense Galactic clouds
\citep*{blg, bergin.cs}.  We also adjust the relative HCN and CO beam
filling factors to account for the likely smaller sizes of the HCN
clouds.  This adjustment generally only affects the resultant
density.

By including the CO/\ico\ integrated intensity ratio at $6\psec8\times
3\psec55$ resolution \citep{n253.ico}, we better constrain the column
density.  Valid solutions are found from the minimum $\chi^2$ (for
$\chi^2 < 1$) with $\phi < 1$, $\phi (1-e^{-\tau}) < 1$ and $T_k >
T_{CO}+2.725$ K.  The parameter $f_1$ is no longer a necessary input.
Given these criteria, we choose the best solution at the lowest $T_k$
with $\chi^2 < 1$.  Given that temperature, the solution for CO column
density per velocity interval and \htwo\ volume density are found from
the minimum $\chi^2$.  Minimizing $T_k$ tends to assign slightly
higher densities, lower column densities, higher beam filling factors
and of course cooler temperatures to the clouds.  Typically a wide
range of valid temperature solutions results, though relative cloud
properties (gradients) are well represented with this procedure.

We do not vary the relative \ico\ and CO beam filling factors, that
is, we assume they are the same based on the clumpy nature of clouds.
Even in massive star formation regions, where isotope-selective 
photodissociation (ISPD) may reduce the abundance of the less
well-shielded molecular species, these lines appear to originate from
essentially the same volume of gas over 100 pc scales
\citep[e. g.,][]{grs},  Further, ISPD does not seem to be important
over large spatial scales in the centers of NGC 253 or other
CO-bright galaxies \citep{umigs}.

\subsection{Maps of Cloud Properties}

Table~\ref{tab.models} lists the model inputs and resultant average
cloud properties.  Models A1--A5 consider only the HCN/CO ratio and
initial beam filling factor, $f_1$.  Models B1--B5 consider the HCN/CO
and CO/\ico\ ratios, and solve for the cloud conditions from the
minimum $\chi^2$ and assuming the minimum valid $T_k$.  Maps of $T_k$,
CO column and \htwo\ volume density ($N_{CO}/\Delta v$ and \nhh), and
beam filling factor are displayed for model solutions A1 and B2 in
Figures~\ref{fig.soln1} and \ref{fig.soln2}.  Figure~\ref{fig.hist}
shows the distributions of cloud properties for model B2.

Varying the model inputs does not greatly affect the morphology seen
in the maps, though the contour values change, as mentioned before
\citep{hcnco}.  Therefore we restrict our discussion to the relative
gas properties and how they vary with position.  However it is worth
mentioning how the model inputs affect the results.  As expected, a
high fractional HCN abundance or high $\phi_{HCN}/\phi_{CO}$ ratio
results in lower \htwo\ density solutions.  Similarly, a high
fractional \ico\ abundance (low [CO]/\ico]) results in lower column
density solutions.  The solution from model B2
(Figures~\ref{fig.soln2} and \ref{fig.hist}) illustrates the effect of
using the minimum temperature.  The northeastern region of NGC 253 is
predicted to have relatively low temperatures ($< 100$ K), but the
beam filling factor there approaches unity ($> 0.6$) and the column
densities are also relatively low ($< 10^{17}$ \coldv), both of which
are expected consequences of underestimating the temperature.  Note
however, that these presumably ``low'' temperature limits average a
warm 50 K.  Indeed, kinetic temperatures in this general region of NGC
253 derived from NH$_3$ observations are predicted to exceed 140 K
\citep{n253.nh3}.  Our derived temperature of over 200 K in the
southwest is twice the estimate from NH$_3$, but considering the
difference in resolution between these datasets, and that both studies
yield limits to $T_k$, the agreement is reasonable.  Given that the
southwestern cloud is also the mid-IR peak and coincident with a
bright cluster of massive stars \citep{keto}, such high temperatures are
not wholly unexpected.

Using the resultant gas properties and the radiative transfer model,
we derive expected CO (\jtw)/(\jon) intensity ratios varying from 1.0
to 1.7 across the central area of the map.  Though difficult to
compare with our map, the \jtw\ observations of \citet{n253.co21}
appear to be consistent with the results.  We expect ratios of 1--1.5
in the central region of the map, and only model B1, with the Galactic
center \ico\ abundance, disagrees in that it predicts CO (\jtw)/(\jon)
$\sim 2$ over most of the map.

The central peaks in temperature, volume and column density are more
or less symmetrically distributed around the kinematic center of NGC
253.  They also flank the area of solid body rotation at the nucleus
and lie perpendicular to the rotation axis of the region
(Figure~\ref{fig.ratv}).  These warm and dense clouds could therefore
very well be part of a nuclear ring as seen in twin-peaks galaxies.
(Note that the very high temperatures derived in models B1--B5 are
still lower limits based on the modeling procedure.)
Figure~\ref{fig.ratv} shows the variation in the orbital velocity and
HCN/CO ratio along the major axis of this supposed ring.  The density
peaks appear at the ends of the central region of solid body rotation
where velocity gradients are lower, and star formation may occur more
readily.

\section{Cloud Kinematics}\label{kin}

We examine the cloud kinematics to estimate the effect of the bar on
driving star formation at the locations of the dense nuclear clouds in
NGC 253.  We primarily search for inner Lindblad resonances (ILRs) and
regions of overlapping orbits and/or low velocity gradients by
analyzing the CO rotation curve and by modeling the bar potential.

\subsection{Rotation Curve}

Because NGC 253 is nearly edge-on, we use a terminal velocity method,
similar to the iteration method of \citet{sofue}, to determine the
rotation curve from the PVD (Figure~\ref{fig.pvd}).  The PVD is
created by first rotating the CO data cube by $51\degr$, the PA of the
major axis \citep{pence}.  To increase the signal-to-noise ratio, we
average the CO spectra within $\pm2''$ of the major axis, which
roughly matches the beam extent.  The rotation curve is estimated at
each point along the major axis from the maximum velocity, corrected
for inclination, at which emission was detected above the $3\sigma$
level in at least two contiguous channels.  From this velocity we
subtract (in quadrature) the half-width values of the line broadening
due to beam smearing, the 10 \kms\ channel width and the turbulence of
the interstellar medium (assumed to be 10 \kms).  The beam smearing is
estimated from the gradient in the rotation curve $\pm 2''$ around
each position.  We then use the CO emission distribution and the
derived rotation curve to recreate the PVD.  The differences between
the observed and modeled PVDs are used to correct the rotation curve,
and a new model PVD is created.  This procedure is repeated and
monitored until convergence.

The high inclination of NGC 253 allows us to derive the radial gas
distribution, $I(r)$, from the observed axial CO distribution,
$I(x)$, unlike the more general technique of \citet{sofue}.  They
assume that the radial distribution is proportional to the observed,
axial distribution.  However, for highly inclined galaxies, foreground
gas contributes significantly to the emission seen towards the central
positions.  Assuming a semi-circular (treating the eastern and western
halves separately), axisymmetric disk geometry,

\begin{equation}
I(x_i) = \sum^{N}_{j=i} I(r_j) \delta A(x_i,r_j),
\end{equation}

\noindent
where $\delta A(x_i,r_j)$ is the section of an annulus between radii
$r_j-\Delta x$ and $r_j+\Delta x$, and major axis positions
$x_i-\Delta x$ and $x_i+\Delta x$, and $\Delta x$ is half the pixel
separation along the major axis.  The radial distribution is then
derived from the outermost observed point,

\begin{equation}
I(r_N) = I(x_N)/\delta A(x_N,r_N),
\end{equation}

\noindent
and back-substitution using

\begin{equation}
I(r_j) = \Big[I(x_j) - \sum^{N}_{i=j+1} I(r_i) \delta A(x_j,r_i)\Big]/
\delta A(x_j,r_j).
\end{equation}

\noindent
The resulting rotation curve is displayed with the PVD in
Figure~\ref{fig.pvd}.  The r.m.s.\ difference between the observed and
reconstructed PVD contours (at $3\sigma$, and 20\%, 30\%, ... 90\%\ of
the peak) is 25 \kms.  The r.m.s.\ difference between the $3\sigma$
level of the PVD, corrected for inclination, and the rotation curve is
11 \kms.  Except near the nucleus, the observed and modeled spectra
agree reasonably well, particularly the line widths and central
velocities; even some double-peaked spectra are reproduced.  Both
sides reach the same ``turnover'' velocity of $\sim 160$ \kms\ at the
same circumnucler distance of $\sim 3''$ (50 pc for $D = 3.4$ Mpc).
The rotation curve is somewhat asymmetric in that the western side
drops more quickly and has a lower final velocity than the eastern
side.  A simple offset in the velocity centroid position and/or
systemic velocity cannot account for this asymmetry.  For example, no
offset in the adopted central position along the major axis results
in a symmetric radial gas distribution, $I(r)$.  A similar asymmetry
is seen in other tracers such as CO \jtw, \ha\ and \htwo\ at
comparable resolution \citep{n253.haspec, n253.co21, n253.h2}, but the
large-scale \hi\ and CO velocities are more symmetric and confirm the
adopted systemic velocity \citep*{n253.hi, sorai}.  Near the nucleus,
the line widths of the modeled spectra underestimate the observed line
widths, and are relatively featureless in comparison.  Apparently the
linear gradient assumed near the nucleus is an oversimplification of
the true gas dynamics \citep[see ][]{n253.h92a}.  It is also likely
that the intrinsic linewidths due to turbulence are higher near the
nucleus.

\subsection{Resonances}

Star formation in nuclear rings is predicted to occur because of
cloud collisions or gravitational instability near ILRs
\citep[e. g.,][ and references therein]{twinpeaks}.  We search for
expected resonances in NGC 253 using the derived rotation curve.  The
bar pattern speed $\Omega_b$ is estimated from the orbital velocity
observed at a radius 20\%\ beyond the ends of the bar and associated
stellar ring \citep{sorai}.  The stellar bar and ring in NGC 253 are
very well defined from recent mid- and near-infrared measurements by
2MASS and the MSX satellite \citep{2mass, msx}, and from \ha\ imaging
\citep{n253.haimg}.  The bar is symmetric about the adopted dynamical
center of the galaxy and terminates at the ring at a projected radius
of $\sim 115''$ (Figure~\ref{fig.2mass}).  We adopt $\Omega_b = 50$
\kms kpc$^{-1}$, the pattern speed derived from CO and \ha\ spectra
\citep{n253.haspec, sorai}.

From the gas angular velocity $\Omega(r) = v(r)/r$ derived from the
rotation curve, $v(r)$, we calculate the epicyclic frequency,

\begin{equation}
\kappa^2 = 4 \Omega \Big(\Omega + \frac{1}{2}R\frac{d\Omega}{dr}\Big).
\end{equation}

\noindent
ILRs are predicted at radii where $\Omega(r)-\kappa(r)/2 = \Omega_b$.
Figure~\ref{fig.omega} shows the angular speed evaluated using each
side of the rotation curve separately, and the weighted average of the
eastern and western rotation curves.  Though no clear ILR is apparent,
$\Omega-\kappa/2$ approaches the pattern speed near radii of $\sim
10$--$15''$ and $25''$.  We will only dicuss the possible inner ILR
since our observations do not effectively probe the outer ILR at
$25''$.  The density peaks from the HCN/CO analysis occur only $\sim
5''$ from the nucleus, implying that they either lie inside this
possible ILR, or are seen in projection.  Figure~\ref{fig.twinp}
compares the HCN/CO and 2MASS $H-K_s$ maps.  The leading dust lanes
end at the central near-infrared features, which are coincident with
the millimeter continuum peak and are aligned with the distribution of
molecular gas density.  The projected PA of the bar minor axis is
$44\degr$, which is close to that of the dense clouds.  In fact, the
dense clouds may lead the minor axis of the bar as seen in the
twin-peaks galaxy NGC 6951 \citep{kohno}.  In NGC 253, the coincidence
and alignment of the density peaks, bar axes and possible ILR imply
that NGC 253 is likely a twin-peaks galaxy, and the bar drives star
formation around its nucleus.

\subsection{Bar Models}

The observed gas distribution and rotation curve are difficult to
reconcile with the typical model describing symmetric stellar orbits
in a bar potential.  It is likely that gas is not uniformly
distributed along the predicted stellar orbits as is seen even in the
center of most face-on, twin-peaks galaxies \citep{twinpeaks, knapen,
sheth}.  Nevertheless, in an attempt to explain the observed gas
motions and density variations as an effect of the bar, we examine the
expected orbits in the center of NGC 253 as in the literature
\citep{n253.cs, sakamoto, das}.  With likely ILRs in NGC 253, we may
adopt the bar potential \citep{bt}

\begin{equation}
\Phi = \onehalf v^2_0 \ln(x^2+y^2/q^2+R^2_c),
\end{equation}

\noindent
where $v_0$, $q$ and $R_c$ are the limiting velocity, bar asymmetry
parameter and core radius, respectively.  Adjusting these parameters
to fit the PVD, we can estimate the gas orbits in this potential
(Figure~\ref{fig.twinp}).  The dense gas near the center of NGC 253
seems to lie near, or slightly lead, the inner $x_2$ orbits, just as
in NGC 6951.  The regions where the $x_1$ and $x_2$ orbits overlap
also coincide with the possible inner ILR.  Fully hydrodynamic models
of gas motions in a bar potential also support this picture that most
star formation occurs near the intersection regions of these different 
orbit families where gas collisions cause shocks
\citep[e. g.,][]{knapen}.

\section{Conclusions}

We present the CO distribution and kinematics in the starburst nucleus
of NGC 253.  The molecular gas is concentrated in several complexes
which are associated with both the near-infrared bar and likely ``twin
peaks'' around the radio nucleus.  The gas densities, temperatures and
column densities are highest in these presumed twin peaks.  The
CO kinematics indicate that the density peaks may also coincide with
an inner ILR just beyond the region of solid body rotation around the
nucleus.  Even without an ILR, the velocity gradient decreases near
the peaks as well, which could also explain the star formation there.

Analysis of the rotation curve and expected gas motions in a bar
potential, compared with the locations of enhanced gas density, lead
us to conclude that the stellar bar drives the star formation in the
starburst nucleus of NGC 253.  This conclusion is supported by the
likely presence of at least one ILR, gas motions consistent with $x_1$
and $x_2$ orbits, leading dust ridges along the infrared bar, and an
apparent twin peaks morphology.

\acknowledgements

The authors thank I. Puerari for his help with the orbit modeling.  We
also thank the anonymous referee for broad suggestions that greatly
improved this paper.  This work was supported in part by grant
211290-5-25875E from the Consejo Nacional de Ciencia y
Tecnolog\'{\i}a, and grant \#60039-32-33 from the Professional Staff
Congress of the City University of New York.  O.Y. is grateful to
INAOE and the Large Millimeter Telescope for support.  This research
made use of the NASA/IPAC Extragalactic Database (NED) which is
operated by the Jet Propulsion Laboratory, California Institue of
Technology, under contract with the National Aeronautics and Space
Administration.  This publication makes use of data products from the
Two Micron All Sky Survey, which is a joint project of the University
of Massachusetts and the Infrared Processing and Analysis
Center/California Institute of Technology, funded by the National
Aeronautics and Space Administration and the National Science
Foundation.

\clearpage

\begin{figure}
\includegraphics[angle=-90,scale=.65]{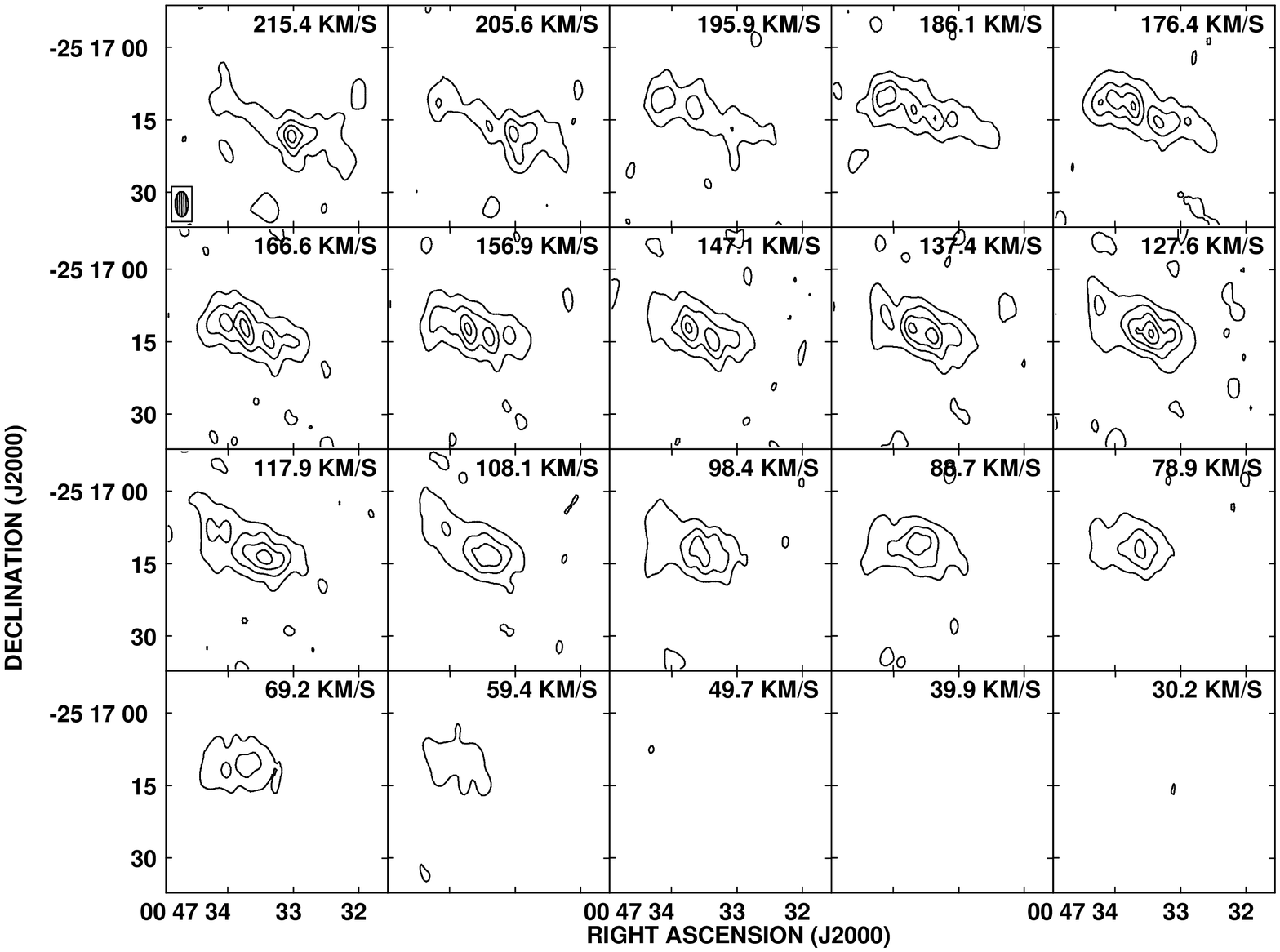}
\end{figure}

\begin{figure}
\includegraphics[angle=90,scale=.85]{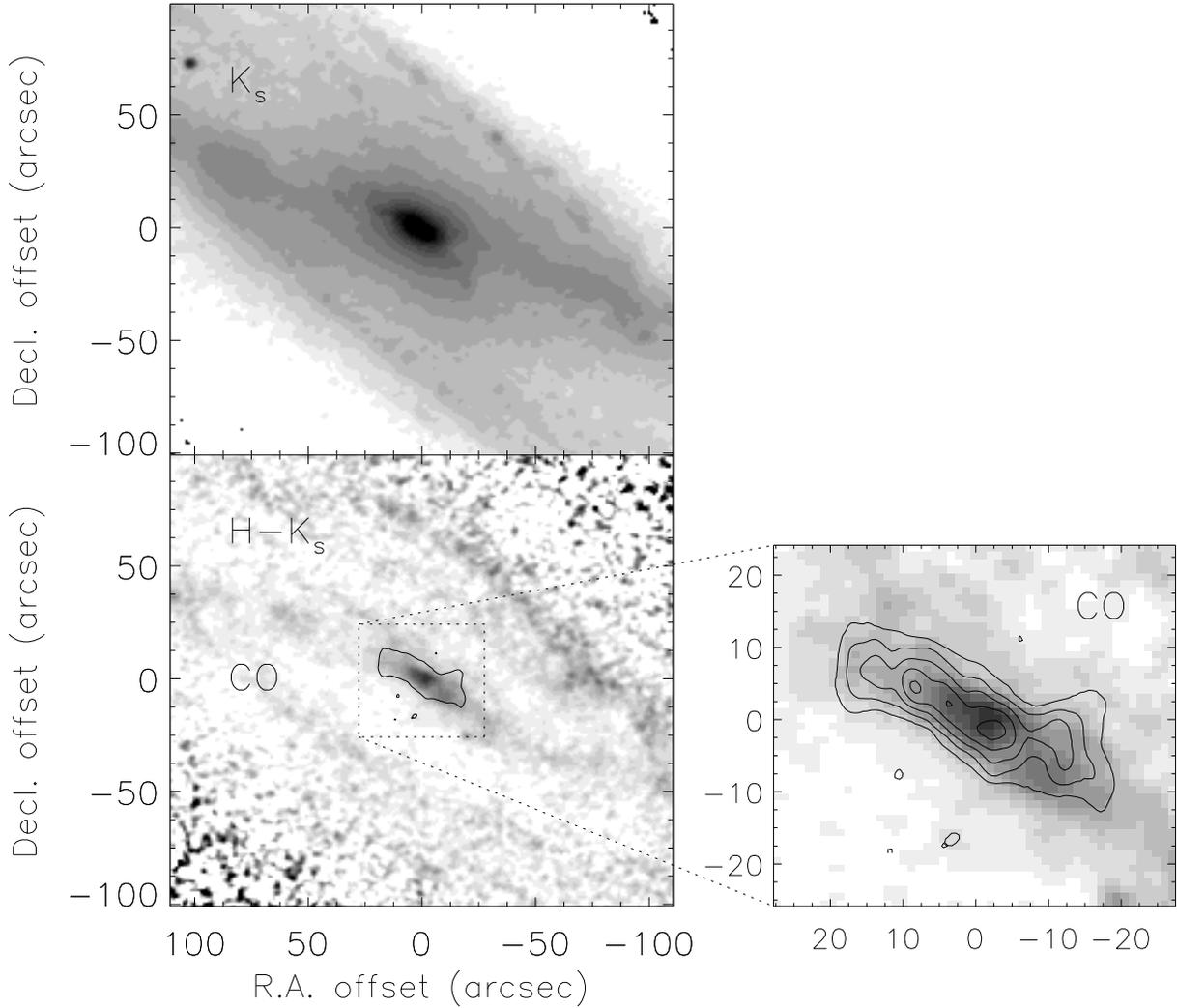}
\caption{(Top) 2MASS $K_s$ image of the central $3\pmin 3$ of NGC
253.  (Bottom left) 2MASS $H-K_s$ image of the same region.  The grey
scale runs from 0.25 to 1.05 mag.  The 10\%\ contour of the CO
emission is indicated.  (Bottom right) $H-K_s$ and CO maps.  CO
contours are as in Figure~\ref{fig.moms}.
\label{fig.2mass}}
\end{figure}

\begin{figure}
\includegraphics[angle=90,scale=.7]{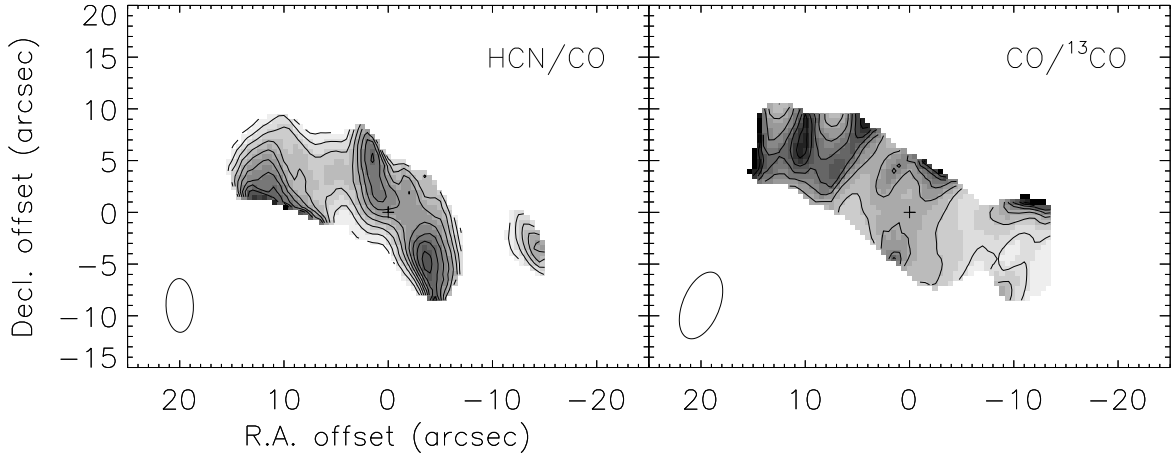}
\caption{(Left) HCN/CO ratio in NGC 253.  Contours are, from white to
black, 0.04 to 0.2 by 0.02.  (Right) CO/\ico\ ratio in NGC 253.
Contours are, from white to black, 5 to 50 by 5.  The location of the
nuclear radio source and the beam size are indicated in each figure.
\label{fig.rats}}
\end{figure}

\begin{figure}
\includegraphics[angle=90,scale=.75]{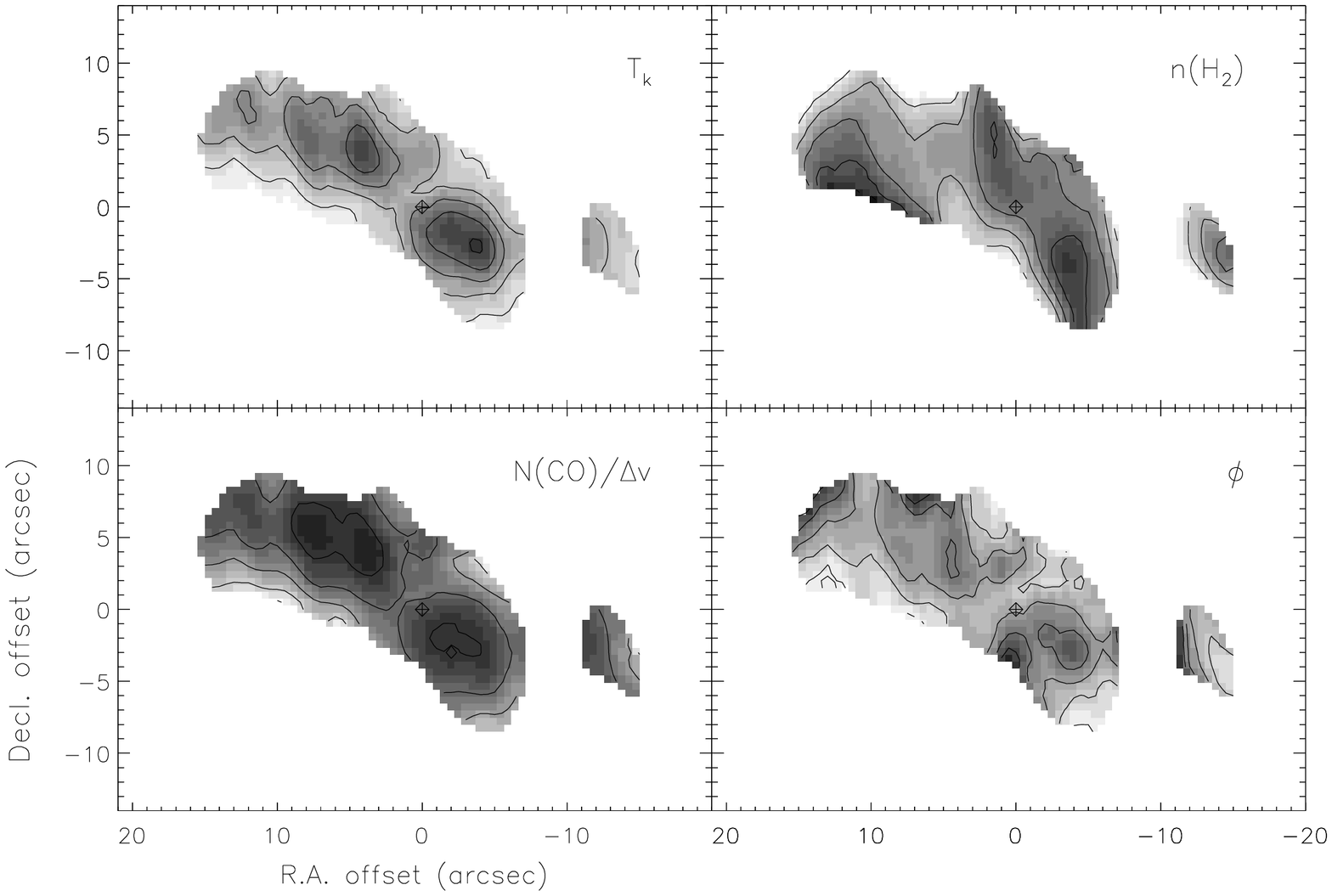}
\caption{Results from model A1 (Table~\ref{tab.models}).  (Clockwise
from upper left) Maps of kinetic temperature, log of \htwo\ density,
CO area beam filling factor, and log of CO column density per velocity
interval in NGC 253.  Contours are 20 to 220 K by 40 K, 2.5 to 5.0
dex by 0.5 dex, 0.12 to 0.20 by 0.02, and 17 to 18 dex by 0.25 dex,
respectively.  The location of the nuclear radio source is indicated.
\label{fig.soln1}}
\end{figure}

\begin{figure}
\includegraphics[angle=90,scale=.75]{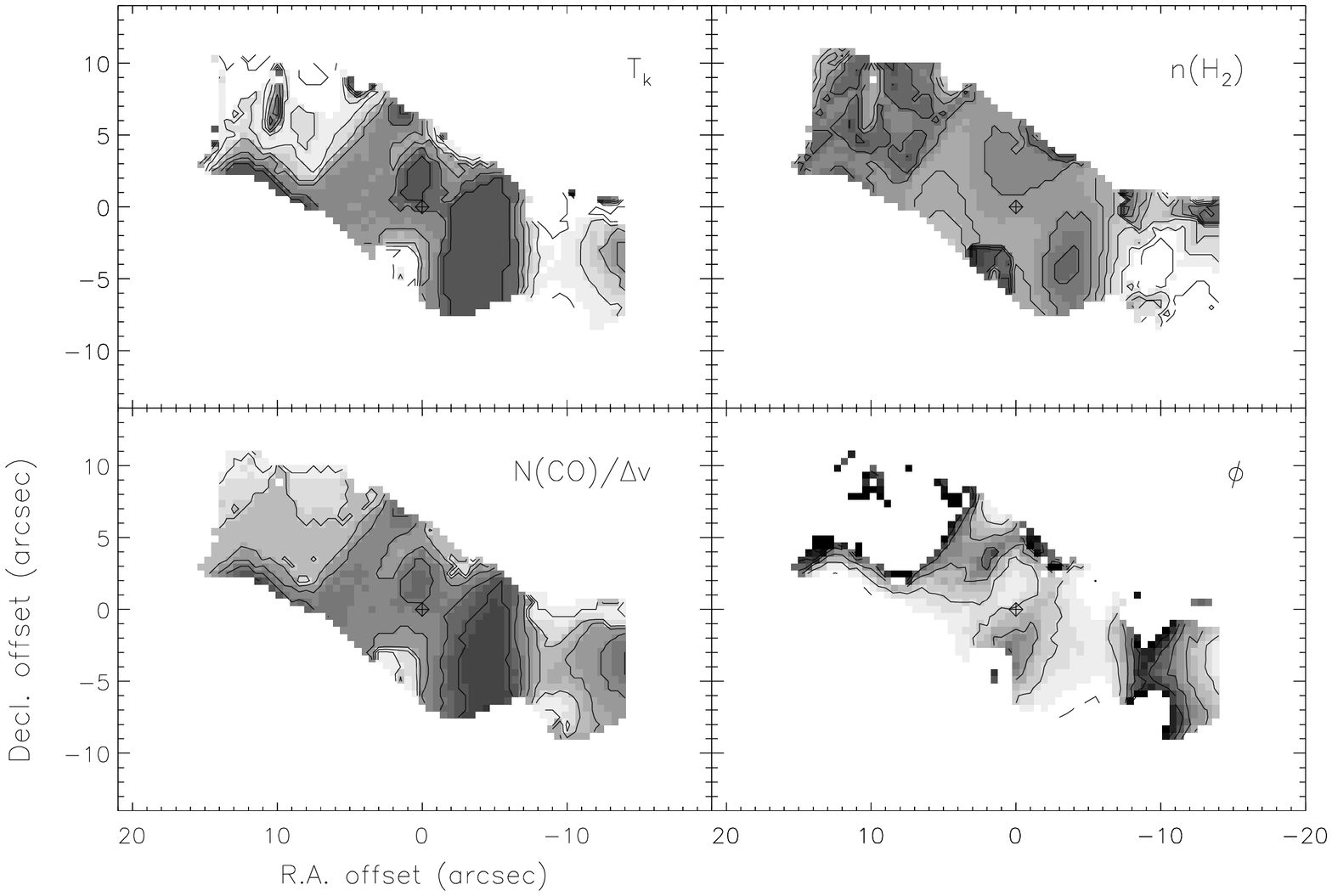}
\caption{Results from model B2 (Table~\ref{tab.models}).  (Clockwise
from upper left) Maps of kinetic temperature, log of \htwo\ density,
CO area beam filling factor, and log of CO column density per velocity
interval in NGC 253.  Contours are 20 to 220 K by 40 K, 3.2 to 4.8
dex by 0.2 dex, 0.05 to 0.55 by 0.1, and 15.75 to 18.25 dex by 0.25
dex, respectively.  The location of the nuclear radio source is
indicated.
\label{fig.soln2}}
\end{figure}

\begin{figure}
\includegraphics[angle=90,scale=.65]{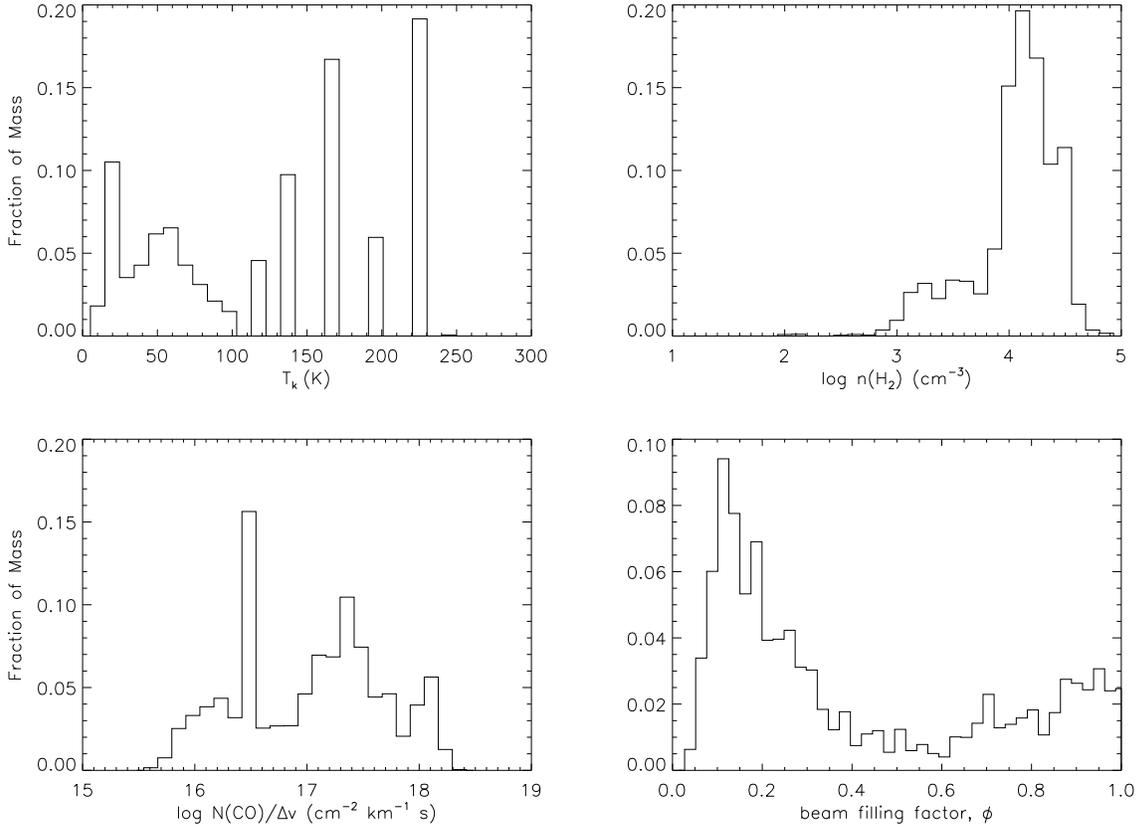}
\caption{Results from model B2 (Table~\ref{tab.models}).  (Clockwise
from upper left) Histograms of kinetic temperature, density, CO area
beam filling factor, and CO column density per velocity interval in
NGC 253.  The fraction of mass is estimated assuming a constant
CO-to-\htwo\ conversion factor.
\label{fig.hist}}
\end{figure}

\begin{figure}
\includegraphics[angle=90]{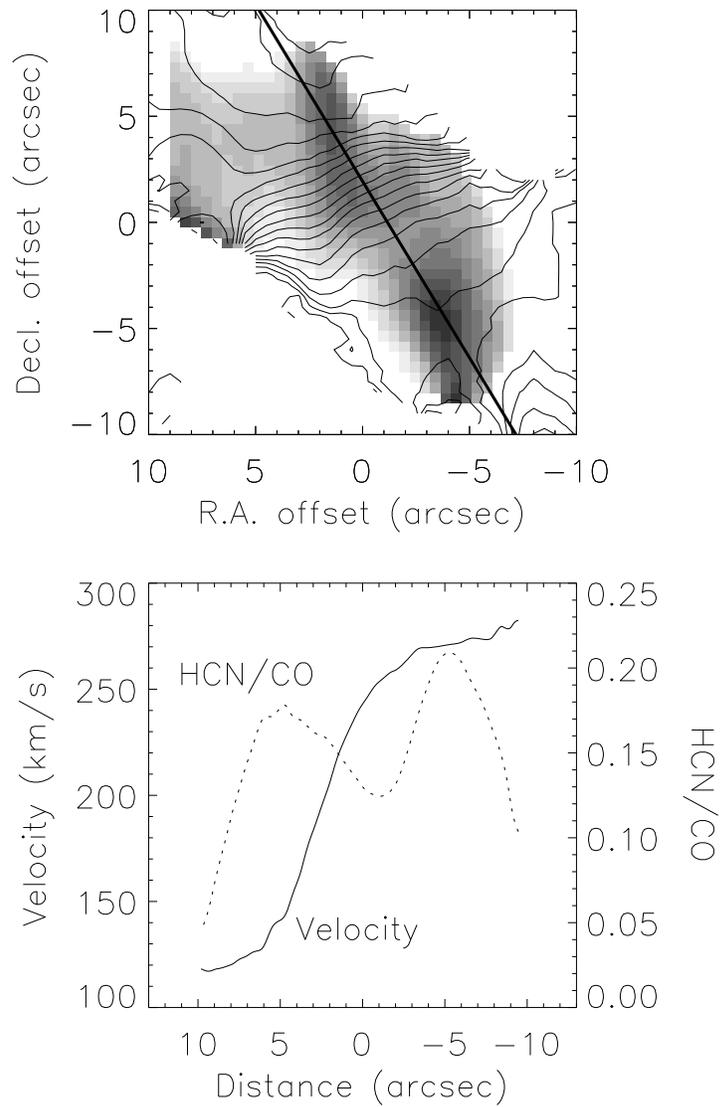}
\caption{(Top) The CO velocity field (contours) over the HCN/CO ratio
(gray scale).  (Bottom) Variation in the ratio and velocity along the
heavy line shown in the top panel.  The line is offset $1''$ northwest
of center and has a position angle of $31\degr$.
\label{fig.ratv}}
\end{figure}

\begin{figure}
\includegraphics[angle=90,scale=.7]{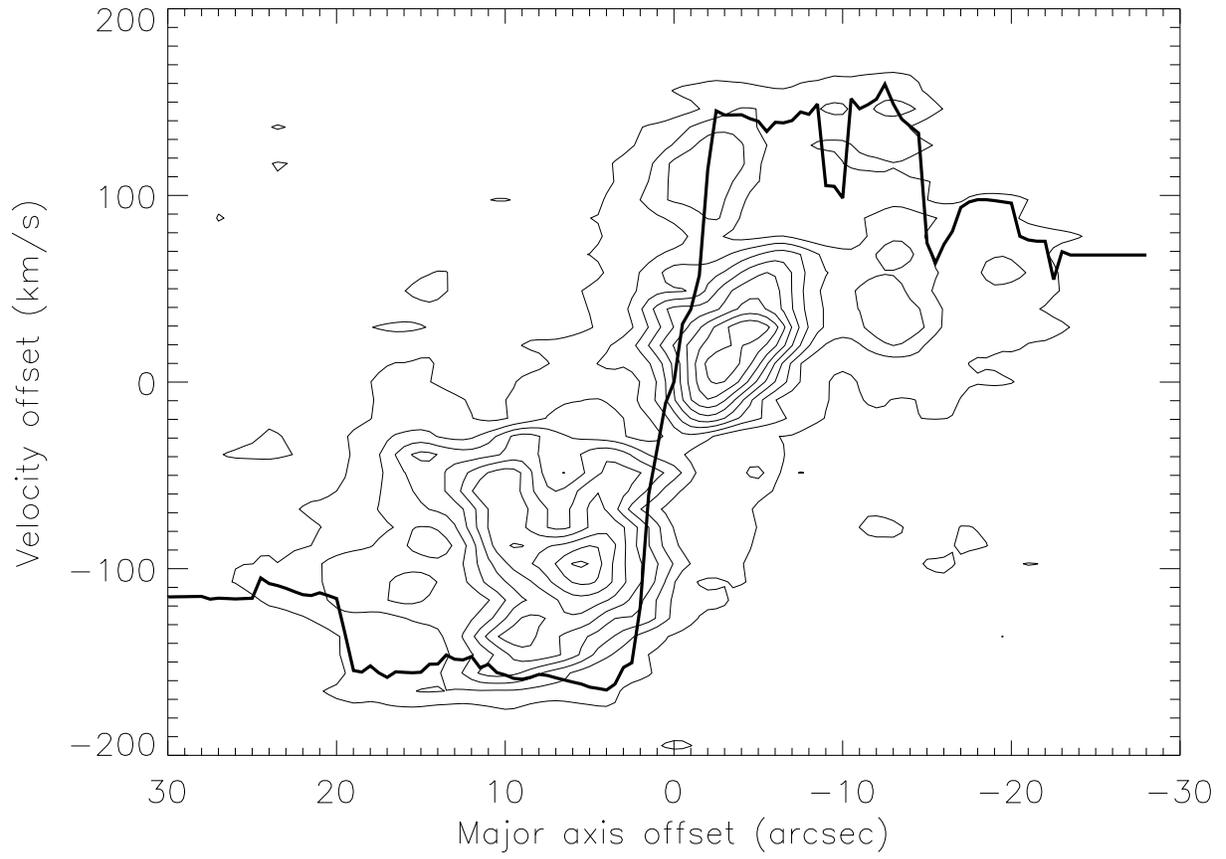}
\caption{Position-velocity diagram of the CO emission (contours) and
the derived rotation curve (heavy line).  The contours are $3\sigma$
and 20\%, 30\%, ... 90\%\ of the peak.
\label{fig.pvd}}
\end{figure}

\begin{figure}
\includegraphics[angle=90,scale=.7]{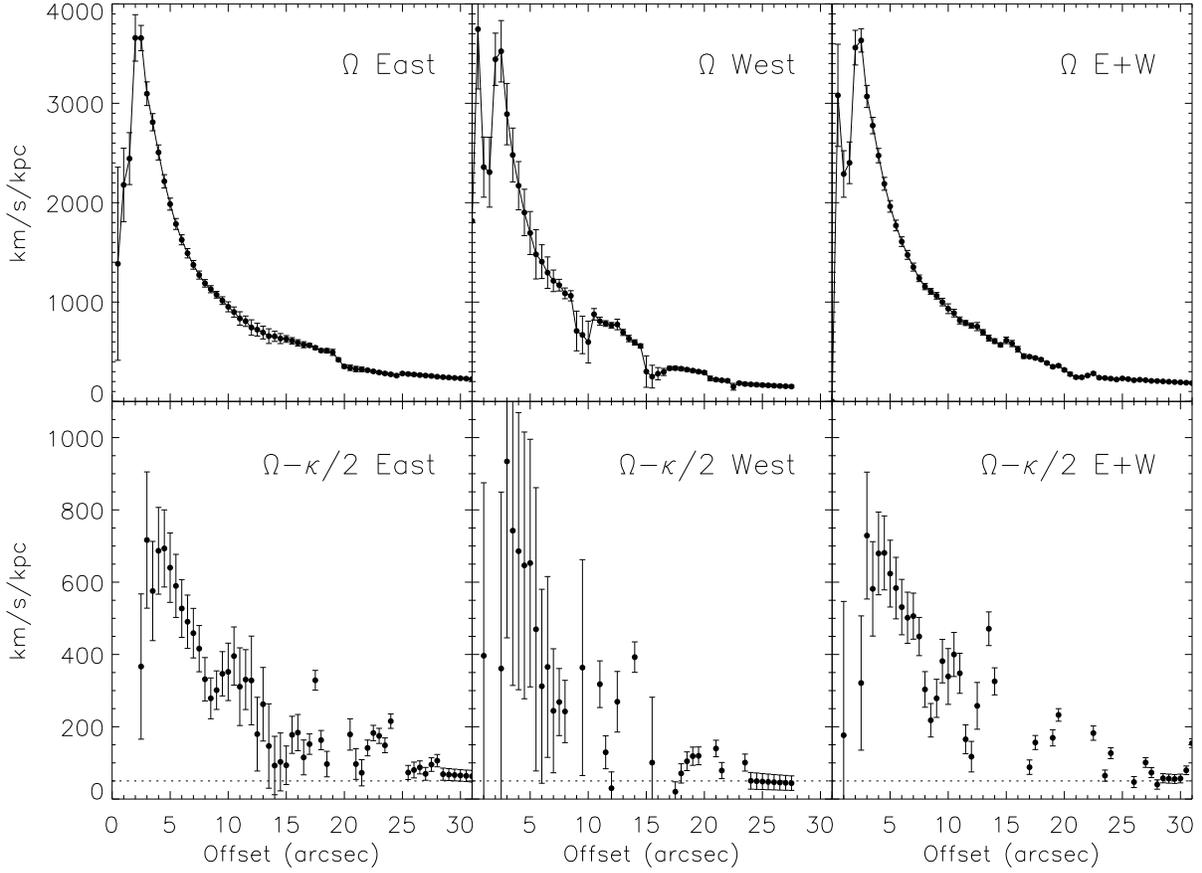}
\caption{(Top) The angular speed $\Omega$ derived from the eastern,
western and combined rotation curves.  (Bottom) Angular speed
($\Omega-\kappa/2$) with the bar pattern speed (dotted line).
\label{fig.omega}}
\end{figure}

\begin{figure}
\includegraphics[angle=90]{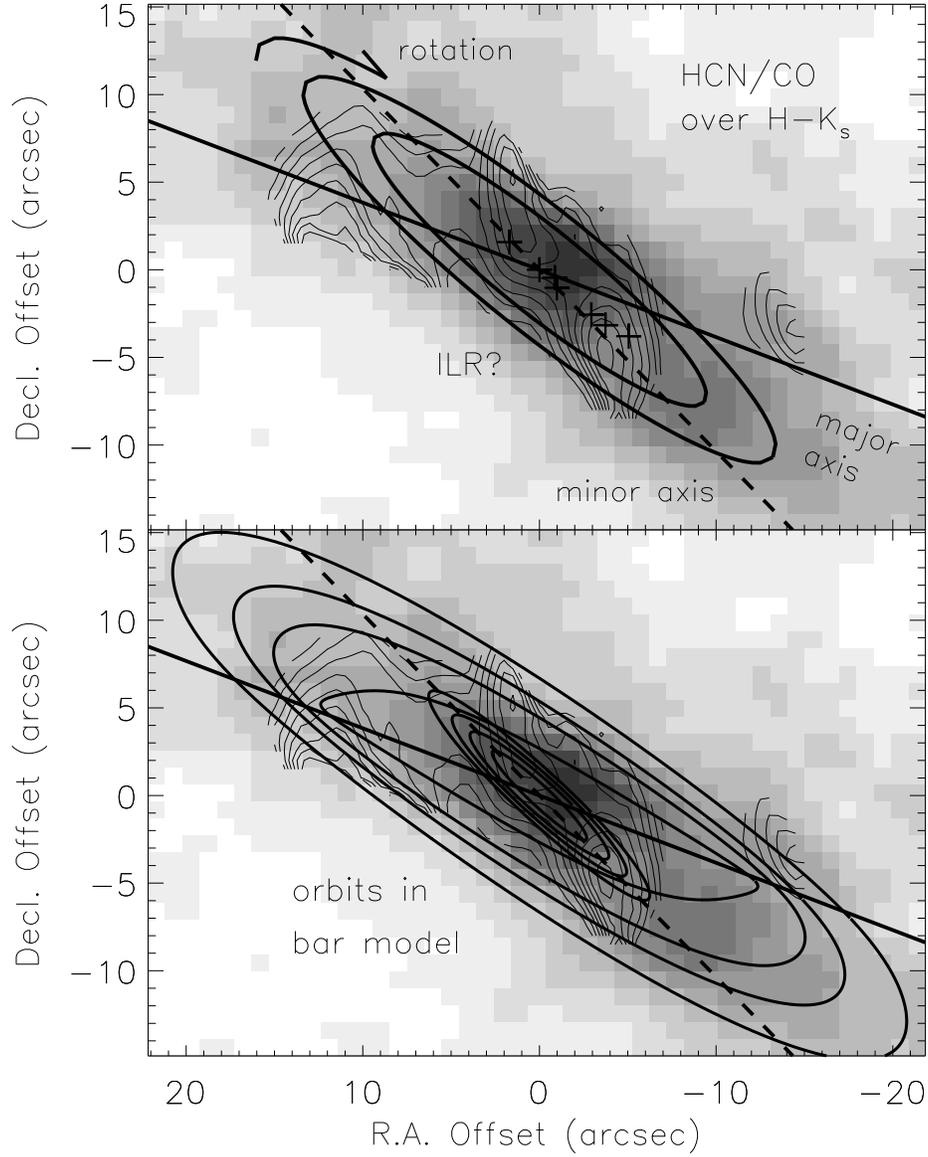}
\caption{(Top) 2MASS $H-K_s$ map (grey scale, from 0.25 to 1.05 mag)
with the HCN/CO intensity ratio (contours).  The bar major and minor
axes, the projected ILR range (12--17$''$), and the thermal radio
sources (crosses) are indicated.  (Bottom) Same as above indicating
the $x_1$ and $x_2$ orbits (outer and inner, respectively) of a bar
model.
\label{fig.twinp}}
\end{figure}

\clearpage

\begin{deluxetable}{l@{}ccccr@{$\,\pm\,$}lr@{$\,\pm\,$}l@{}r@{$\,\pm\,$}lr@{$\,\pm\,$}l}
\tablewidth{0pt}
\tablecaption{Modeling Inputs and Results$^a$
\label{tab.models}}
\tablehead{
Model &
$f_1$ & $\frac{\phi_{HCN}}{\phi_{CO}}$ & [HCN] & 
$\frac{[{\rm CO}]}{[\ico]}$ & 
\multicolumn{2}{c}{$\langle T_k \rangle$} &
\multicolumn{2}{c}
{$\langle\log(N_{CO}/\Delta v)\rangle$}
& \multicolumn{2}{c}{$\langle\log n\rangle$} & 
\multicolumn{2}{c}{$\langle\phi\rangle$} \\
& & & \ee{8} & & \multicolumn{2}{c}{(K)} & \multicolumn{2}{c}{(\coldv)} & 
\multicolumn{2}{c}{(\percc)} &\multicolumn{2}{c}{}
}
\startdata
A1 & 0.1  & 1.0 & 2 & \nodata & 130 & 40 & 17.8 & 0.2 & 3.8 & 0.5 & 0.16 & 0.02\\
A2 & 0.1  & 0.5 & 2 & \nodata & 130 & 40 & 17.8 & 0.2 & 4.7 & 0.6 & 0.16 & 0.02\\
A3 & 0.2  & 1.0 & 2 & \nodata &  70 & 20 & 17.5 & 0.2 & 3.9 & 0.4 & 0.27 & 0.02\\
A4 & 0.2  & 0.5 & 2 & \nodata &  70 & 20 & 17.4 & 0.2 & 4.7 & 0.5 & 0.26 & 0.02\\
A5 & \phn0.05 & 1.0 & 2 & \nodata & 240 & 60 & 18.1 & 0.1 & 3.8 & 0.5 & 0.10 & 0.02\\
B1 & \nodata & 1.0 & 2 & 30 &  90 & 60 & 16.7 & 0.5 & 4.2 & 0.3 & 0.57 & 0.28\\
B2 & \nodata & 1.0 & 2 & 50 & 120 & 70 & 17.0 & 0.6 & 4.0 & 0.4 & 0.41 & 0.31\\
B3 & \nodata & 0.2 & 2 & 50 & 190 & 50 & 17.9 & 0.7 & 6.0 & 1.1 & 0.17 & 0.24\\
B4 & \nodata & 1.0 & \phn\phd 0.7 & 50 & 130 & 90 & 17.3 & 0.8 & 5.2 & 1.3 & 0.40 & 0.37\\
B5 & \nodata & 1.0 & 6 & 50 &  70 & 40 & 16.6 & 0.4 & 3.6 & 0.5 & 0.58 & 0.28\\
\enddata
\tablenotetext{a}{Brackets denote mass-weighted averages assuming a
constant $\htwo$-to-CO conversion factor.  Values are listed with
$1\sigma$ dispersions.}
\end{deluxetable}

\end{document}